**Title**

# Harnessing dislocation motion using an electric field

**Author list**

Mingqiang Li[1,2], Yidi Shen[3], Kun Luo[3], Qi An[3*], Peng Gao[2,5*], Penghao Xiao[4*], Yu Zou[1*]

**Affiliations**

[1]Department of Materials Science and Engineering, University of Toronto, Toronto, ON M5S 3E4, Canada

[2]Electron Microscopy Laboratory, and International Center for Quantum Materials, School of Physics, Peking University, Beijing 100871, China

[3]Department of Materials Science and Engineering, Iowa State University, Ames, IA 50011, United States of America

[4]Department of Physics and Atmospheric Science, Dalhousie University, Halifax, Nova Scotia B3H 4R2, Canada

[5]Interdisciplinary Institute of Light-Element Quantum Materials and Research Center for Light-Element Advanced Materials, Peking University, Beijing 100871, China

*Corresponding authors. Emails: qan@iastate.edu; p-gao@pku.edu.cn; penghao.xiao@dal.ca; mse.zou@utoronto.ca

**Abstract**

Dislocation motion, an important mechanism underlying crystal plasticity, is critical for the hardening, processing, and application of a wide range of structural and functional materials. For decades, the movement of dislocations has been widely observed in crystalline solids under mechanical loading. However, the goal of manipulating dislocation motion via a non-mechanical field alone remains elusive. Here we present real-time observations of dislocation motion controlled *solely* by using an external electric field in single-crystalline zinc sulfide (ZnS) – the dislocations can move back and forth depending on the direction of the electric field. We reveal the nonstoichiometric nature of dislocation cores and determine their charge characteristics. Both negatively and positively charged dislocations are directly resolved, and their glide barriers decrease under an electric field, explaining the experimental observations. This study provides direct evidence of dislocation dynamics controlled by a non-mechanical stimulus and opens up the possibility of modulating dislocation-related properties.

**Main text**

The modulation of mechanical and functional properties of crystalline solids based on dislocation mobility has been a central issue in materials research since the introduction of dislocation theory in the 1930s[1, 2, 3]. For most crystalline solids, plastic deformation is achieved by dislocation motion. Due to their highly mobile dislocations, ductile metals can be deformed into final products through compression, tension, rolling, and forging[4]. In contrast, ionic and covalent crystals generally suffer from poor dislocation mobility, rendering them too brittle to process using mechanical methods, and thus they are often unsuited to a broad range of manufacturing techniques[5, 6]. The movement of dislocations also substantially affects the lifetime of semiconductor devices[7], crystal growth[8], solid phase transition[9], and stress-induced amorphization of crystals[10, 11].

Dislocation motion is generally associated with mechanical stresses, and its related dislocation dynamics have been thoroughly studied[12, 13, 14, 15]. Furthermore, extensive studies have been carried out to manipulate dislocation mobility by applying both mechanical stress and an additional stimulus. The results of these studies show that, under mechanical loading, flow stresses can be influenced by an electric field[16, 17, 18, 19] in a phenomenon known as electroplasticity[16, 20], or by illumination[21], in a phenomenon known as photoplasticity[22, 23, 24]. For example, the dislocation mobility of metallic, ionic, and covalent crystals may be enhanced by an electric field through Joule heating, electron wind force, charged dislocations, and other hypothetical effects[25, 26, 27]. The principal driving force of dislocation motion is, however, still mechanical stress, and this limits the processing approaches and engineering applications of many crystalline materials. The dislocation motion controlled by a non-mechanical stimulus alone is rarely expected.

Early studies on alkali halides under an electric field reported optical microscopic observation of new etching pits on alkali halides surfaces, suggesting possible electrical response of dislocations[28, 29, 30]. However, to date, real-time observation of dislocation motion under an electric field has rarely been reported and related dislocation dynamics are largely unknown. Due to recent advances, in situ transmission electron microscopy (TEM) has provided opportunities for visualizing crystalline defects and investigating dislocation motion in the presence of external stimuli[14, 31, 32, 33, 34]. For example, the movement of dislocations driven by mechanical stresses has been ubiquitously visualized in various crystals[14, 35]. The generation and jamming of dislocation clouds by electron wind force was investigated in the phase transition materials $Ge_2Sb_2Te_5$[36]. Also, random dislocation motion induced by electron beam irradiation was observed in graphene and GaAs[37, 38]. Despite these efforts, the direct probing of the dislocation motion governed by a non-mechanical stimulus has remained out of reach. In this study, the use of in situ TEM made it possible to observe that the motion of dislocations in ZnS is driven by an external electric field. This phenomenon can be explained via the atomic-scale characterization of dislocation cores and density functional theory (DFT) calculations of the dislocation glide barriers.

Fig. 1 illustrates how an individual dislocation line in a single-crystalline ZnS sample (Extended Data Fig. 1) can be subjected to an external electric field and observed in situ using TEM. As shown in Fig. 1a and Supplementary Fig. 1, a tungsten tip was used to apply a variable electric field to the ZnS sample. Initial in situ TEM testing was performed on a single 30° partial dislocation (Supplementary Fig. 2). As shown in Supplementary Video 1, the dislocation starts to move when the voltage reaches ~50 V (an electric field of ~$2.4 \times 10^7$ V m$^{-1}$); the movement depends on the direction of the applied electric field. Under a positive voltage bias (Fig. 1c), the dislocation moves away from the tip. When an opposite electric field is applied, the direction of the dislocation motion is reversed, moving towards the tip (Fig. 1d). In this experiment, the dislocation moves in a reversible manner within a range of ~82.1 nm, reaching a maximum velocity of ~306.4 nm s$^{-1}$ (Supplementary Fig. 3).

Fig. 2 shows that dislocation mobility under an electric field depends on the type of dislocation. Fig. 2a displays the initial positions of two 30° partial dislocations (Dislocations B and D, labelled at the bottom of Fig. 2a) and three 90° partial dislocations (Dislocations A, C, and E) (see Extended Data Fig. 2 for the identification of these dislocations). Dislocations B and D start to move to the left when the applied voltage is above ~100 V (Supplementary Video 2), and they continue to move (Fig. 2b) until they are adjacent to Dislocations A and C (Fig. 2c). As the voltage decreases to 0 V from the peak voltage value of 149 V, each of the five dislocations remains in the same position (Fig. 2d). When an opposite electric field is applied (Fig. 2e and f), the movements of Dislocations B and D are reversed. Dislocations A, C, and E remain motionless through all voltage changes. The measured displacement and applied voltage of the two marked points on Dislocations B and D are plotted as functions of time (Supplementary Fig. 4 and Fig. 2g, respectively). The maximum displacement of the marked point on Dislocation B is about 292.6 nm with velocity ranges from 0-623.5 nm s$^{-1}$ (Supplementary Fig. 4). As shown in Fig. 2h, Dislocations B and D start to move when the voltage exceeds a threshold of ~100 V, corresponding to an electric field of ~$5.8 \times 10^6$ V m$^{-1}$ applied to the dislocations (Supplementary Fig. 5 and Supplementary Table 1). The measured displacements of Dislocations B and D do not exhibit a linear relationship as a function of the applied voltage due to the pinning-depinning (Extended Data Fig. 3) and kink mechanisms during motion (Extended Data Fig. 4). Due to relatively high Peierls barriers for dislocation motion in covalent crystals, their dislocations typically move by a kink mechanism[39], which is in good agreement with our experimental observation (Extended Data Fig. 4). Furthermore, it is observed that the 30° dislocations in Fig. 2 move approximately against the direction of the applied electric field, while the 30° dislocation in Fig. 1 moves approximately along the direction of the electric field, implying opposite electrical responses. The phenomenon of electric-field-driven dislocation motion has been repeatedly observed in our experiments (Supplementary Fig. 6).

To explain the effect of an electric field on the dislocations in ZnS, the atomic structures and electronic structures of dislocation cores were characterized using atomic-scale imaging and DFT calculations. Fig. 3a and b are atomically resolved high-angle annular dark field (HAADF) images of 30° S and 30° Zn dislocation cores with S and Zn terminated elements, respectively. Both cations and anions at the dislocation cores are resolved in these $Z$-contrast HAADF images (a brighter atomic column suggests a larger atomic number $Z$[40]). Fig. 3c and d show the Zn-S bond length corresponding to the circles in Fig. 3a and b, respectively. The stacking faults and dislocation cores can be located based on irregularities in the arrangements of atoms compared to the perfect lattice. The dislocation cores modify the electronic structure by introducing additional states into the crystalline band structure[41, 42]. The density of states (DOS) of the 30° S core has a local empty state (acceptor) close to the valence band maximum, which tends to attract extra electrons from the bulk Fermi Sea (Supplementary Fig. 7). In contrast, the 30° Zn core has a local occupied state (donor) close to the conduction band minimum, which tends to lose electrons and become positively charged. The energies of dislocations in neutral and charged states were calculated as functions of the Fermi level ($E_F$) within the band gap (Supplementary Fig. 8). Compared with the neutral states, the negatively charged 30° S core and positively charged 30° Zn core have lower calculated energies within almost the whole range of the band gap. In other words, charged partial dislocations are energetically favorable in ZnS. Fig. 3 and Extended Data Fig. 5 show the net charge distributions of 30° cores and 90° cores, respectively. The blue clouds in Fig. 3e represent the additional electrons ($e^-$) localized around the 30° S core, while the red clouds in Fig. 3f represent the additional holes ($h^+$) around the 30° Zn core. The charged nature of the dislocations makes possible the manipulation of the dislocation movement using an external electric field through electrostatic interaction. Furthermore, the opposite charge states of the 30° Zn core and 30° S core, i.e., the Zn core is positively charged and the 30° S core is negatively charged, may also explain the opposite movement directions of the dislocations under electric fields with the same direction, as shown in Figs. 1 and 2. The dislocation in Fig. 1 is determined to be a positively charged 30° Zn-core dislocation, because the dislocation moves along the same direction as the electric field. Dislocations B and D in Fig. 2 are determined to be negatively charged 30° S-core dislocations, because these dislocations move along the opposite direction from the electric field. We have also observed different dislocations in the same specimen moving in opposite directions under the same electric field (Supplementary Fig. 9), further verifying our charged dislocation analysis.

To further understand the mechanisms of the observed electric field controlled dislocation motion, the minimum energy paths (MEPs) of dislocation glide were analyzed using the nudged elastic band (NEB) method[43, 44]. The MEP is the most plausible way for a dislocation to move. Fig. 4a-c shows the initial, transition, and final states of the glide process of a 30° S partial dislocation as viewed along the [110] and [111] directions. The thickness of our atomic models is ~0.385 nm, which is the length of dislocation lines in our calculation. The glide is accompanied by the bending of a Zn-S-Zn bond in the core region (Fig. 4b). Eventually, the dislocation moves from Position 0 to Position 1 (Fig. 4c). The glide processes of the 30° Zn core, 90° S core, and 90° Zn core were analyzed using the same method (Supplementary Fig. 10). As shown in Fig. 4d, the glide barriers ($\Delta E$) of the 90° partials are higher than those of the 30° S partials, explaining that the 90° partials are sessile (Fig. 2). As discussed above, these nonstoichiometric dislocations tend to be charged by exchanging electrons with the bulk; thus, the glide barrier can be altered by applying an electric field. The enthalpy of a charged system under an electric field $\boldsymbol{\epsilon}$ is modified, to first-order approximation, by the addition of a work term ($W$):

$$W = \sum_i q_i \mathbf{r}_i \boldsymbol{\epsilon} \qquad\qquad (1)$$

where $q_i$ and $\mathbf{r}_i$ are the charge and Cartesian coordinates of atom $i$, respectively. Fig. 4e shows that the barrier of the 30° S core decreases as the strength of the applied electric field increases: the barrier decreases from 0.413 eV nm$^{-1}$ to 0.093 eV nm$^{-1}$ when an electric field of 3 V nm$^{-1}$ is applied, given a net charge of 0.5 e$^-$. Such a small energy barrier (comparable to k$T$ at 300 K, ~0.026 eV) and the overall downhill trend suggest a higher likelihood that the 30° S partial dislocations move in this direction under the electric field. As shown in the MEPs of other dislocations (Extended Data Fig. 6 and Fig. 4f), the glide barriers of all dislocations tend to decrease under an electric field, elucidating the mechanisms of electric field controlled dislocation motion from an energetic perspective. The DFT results show that the glide barrier of the 30° Zn-core dislocation is higher than that of the 30° S-core dislocation, which agrees well with our experiments. The estimated critical electric field required to drive the motion of the 30° Zn-core dislocation in Fig. 1 is ~2.4 × 10$^7$ V m$^{-1}$; while that to drive the 30° S-core dislocations in Fig. 2 is ~5.8 × 10$^6$ V m$^{-1}$. The estimated electrostatic stress on the charged dislocations is in the same order of magnitude as the calculated critical resolved shear stress for the curved dislocations (Supplementary Fig. 11 and Supplementary Table 2). The electric-field-driven dislocation motion may be temperature dependent, due to the thermal fluctuation of atoms and other effects[5, 45]. In general, the Peierls barriers of dislocations decrease with increasing temperature[46], suggesting that a lower electric field is required to trigger dislocation motion at elevated temperatures.

Previous studies have reported that electrically assisted dislocation motion may result from Joule heating[26], electron wind force[25], and charged dislocations[21, 47]. Our observation of dislocations moving back and forth (Fig. 1) eliminates the Joule heating mechanism because the thermal stress arising from Joule heating is independent of the direction of the applied electric field. Our results also rule out the possibility that electron wind force plays a dominant role because the different types of dislocations move in opposite directions under the same voltage (Figs. 1 and 2). Moreover, although irradiation by an electron beam may enhance dislocation mobility due to thermal fluctuation and other irradiation effects[37, 38], this motion is reported to be random, and, in our study (Supplementary Fig. 12), no reversible motion was observed under electron beam irradiation (dose ~30 e Å$^{-2}$ s$^{-1}$). Even when the electron beam is off, we still observe that the positions of the 30° dislocations are changed under an electric field (Extended Data Fig. 7 and Supplementary Video 3). Both atomic-scale characterization and DFT calculations revealed the nonstoichiometric and charged nature of dislocations in ZnS, which explains the observed dislocation motion controlled by an electric field. Charged dislocations exist in a wide range of ionic crystals and semiconductors[21, 48, 49, 50]. Thus, our method could be expanded to other ionic and covalent crystals, especially most II-VI compounds which exhibit relatively lower hardness and lower dislocation glide barriers than elemental and III-V semiconductors[21, 51].

We demonstrate that the dislocation glide barriers can be modified by altering the dislocation charge states and by applying external electric fields, enabling the plastic deformation of crystalline materials by an electric field. Our work may offer an alternative strategy for reducing the dislocation density of semiconductors and insulators. The ability to drive an individual dislocation using an electric field would allow dislocation lines to act as one-dimensional channels for tuning the properties of semiconductors and insulators. Under such circumstances, driving dislocation motion without contacting samples can be achieved (Supplementary Fig. 13). The real-time observation of dislocation motion in this study may provide a useful method for studying the dynamic behavior of crystal defects under an electric field, opening the door to new opportunities for dislocation engineering in materials.

## Acknowledgements

M.L. and Y.Z. acknowledge the financial support from NSERC-Discovery Grant #RGPIN-2018–05731 and CFI-John R. Evans Leaders Fund (JELF) #38044 and # 43597 and Dean's Spark Assistant Professorship from the University of Toronto. Y.S. K.L. and Q.A. were supported by NSF (CMMI-2019459). P.X. acknowledges the support of NSERC-Discovery Grant # RGPIN-2022-02969 and ACENET and the Digital Research Alliance of Canada. P.G. acknowledge the support from the National Key R&D program of China (2019YFA0708200), the National Natural Science Foundation of China (Grant Nos. 52125307 and 52021006), the "2011 Program" from the Peking-Tsinghua-IOP Collaborative Innovation Center of Quantum Matter, and Hefei National Laboratory. We thank J. Xu, J. Zhang and X. Ma at Electron Microscopy Laboratory of Peking University for the support on FIB and TEM; P. Gu and Y. Ye at Peking University for electrical measurements; and C. Wang at Shengzhen University and R. Shi at Peking University for helpful discussion and suggestions. Y.Z. acknowledges R. Spolenak at ETH Zurich for his inspiring initial discussion.

## Author contributions

Y.Z. initiated the idea, developed the research theme, supervised the project and prepared the manuscript outline. M.L. designed and carried out microfabrication and experiments, analyzed the data, and prepared the manuscript draft. P.G. provided the experimental support and supervision; P.X. provided the guidance of DFT calculations and discussion; Y.S. and K.L. carried out DFT calculations under the supervision of Q.A. and P.X. All authors contributed to this work through useful discussion, revision, and comments to the manuscript.

## Competing interests

The authors declare no competing interests.

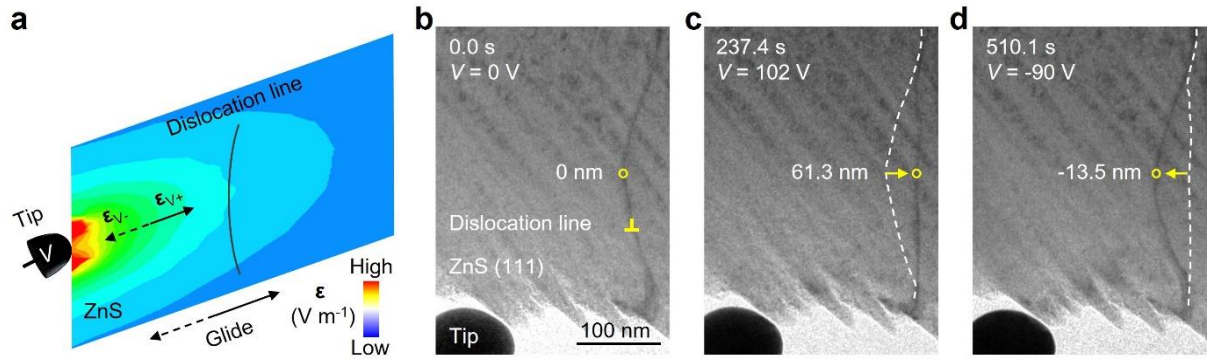

**Fig. 1. Dislocation motion driven by an external electric field. a**, Schematic of the experimental set-up in TEM. A voltage $V$ is applied to a ZnS sample through a tungsten tip. **b-d**, Chronological bright-field TEM images show the positions of a dislocation as a variable voltage is applied. The initial position of the dislocation when $V = 0$ (**b**). The marked point (the yellow circle) moves 61.3 nm to the right (away from the tip) when the voltage increases from 0 to 102 V (**c**). The marked point moves 13.5 nm to the left (towards the tip) when the voltage decreases to -90 V (**d**). The dashed lines in **c** and **d** indicate the positions before dislocation motion. The yellow arrows show the direction of movement of the dislocation.

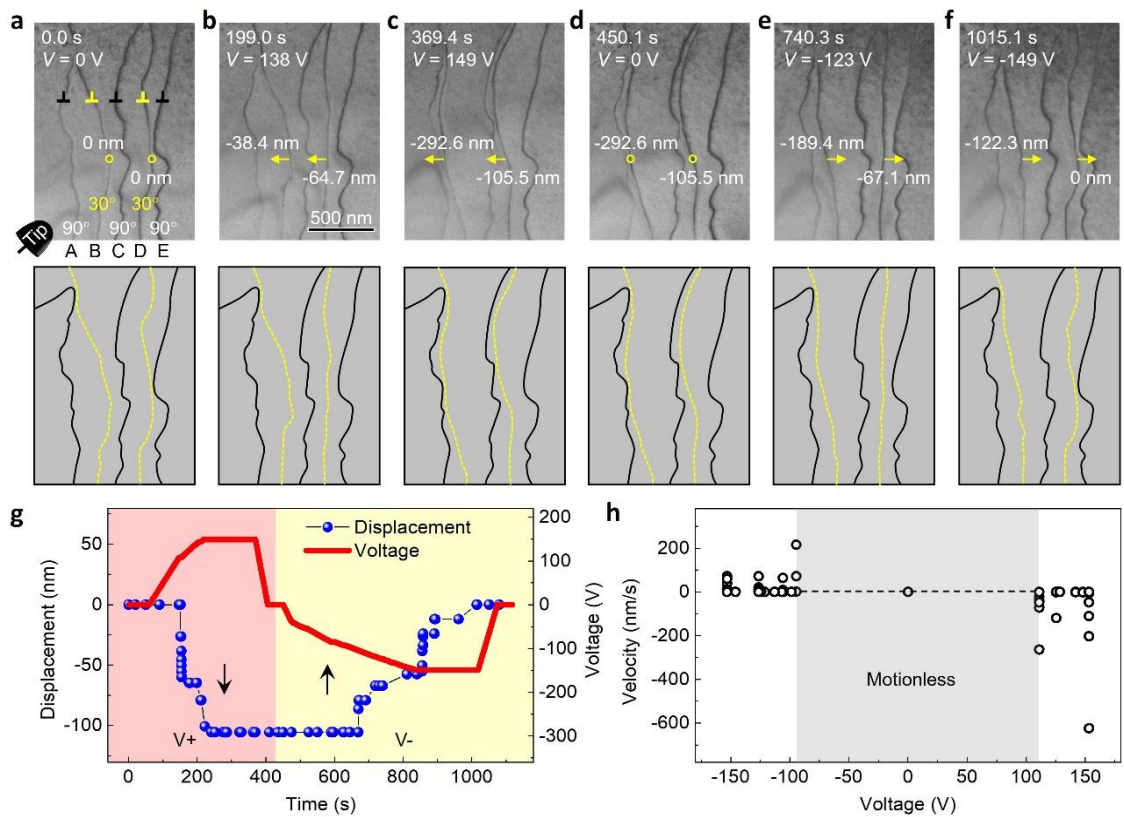

**Fig. 2. The distinct mobility of 30° and 90° partial dislocations under an electric field. a-f**, Chronological bright-field TEM images and the corresponding schematics show the positions of partial dislocations during the experiment: two 30° (B and D, represented as yellow dashed lines) and three 90° (A, C, and E, represented as solid black lines). The yellow arrows indicate that the marked points on Dislocations B and D move to the left under a positive voltage (**b** and **c**) and to the right under a negative voltage (**e** and **f**). Dislocations A, C, and E remain in their initial positions during the entire experiment. **g**, Plots of the applied voltage and displacement of the marked point on Dislocation D as functions of time. The arrows indicate the opposite directions of the dislocation motion under a positive voltage (the red region) and a negative voltage (the yellow region). **h**, The measured velocity of the two marked points in **a** as functions of the applied voltage. The voltage range in which the dislocations are motionless is indicated in grey.

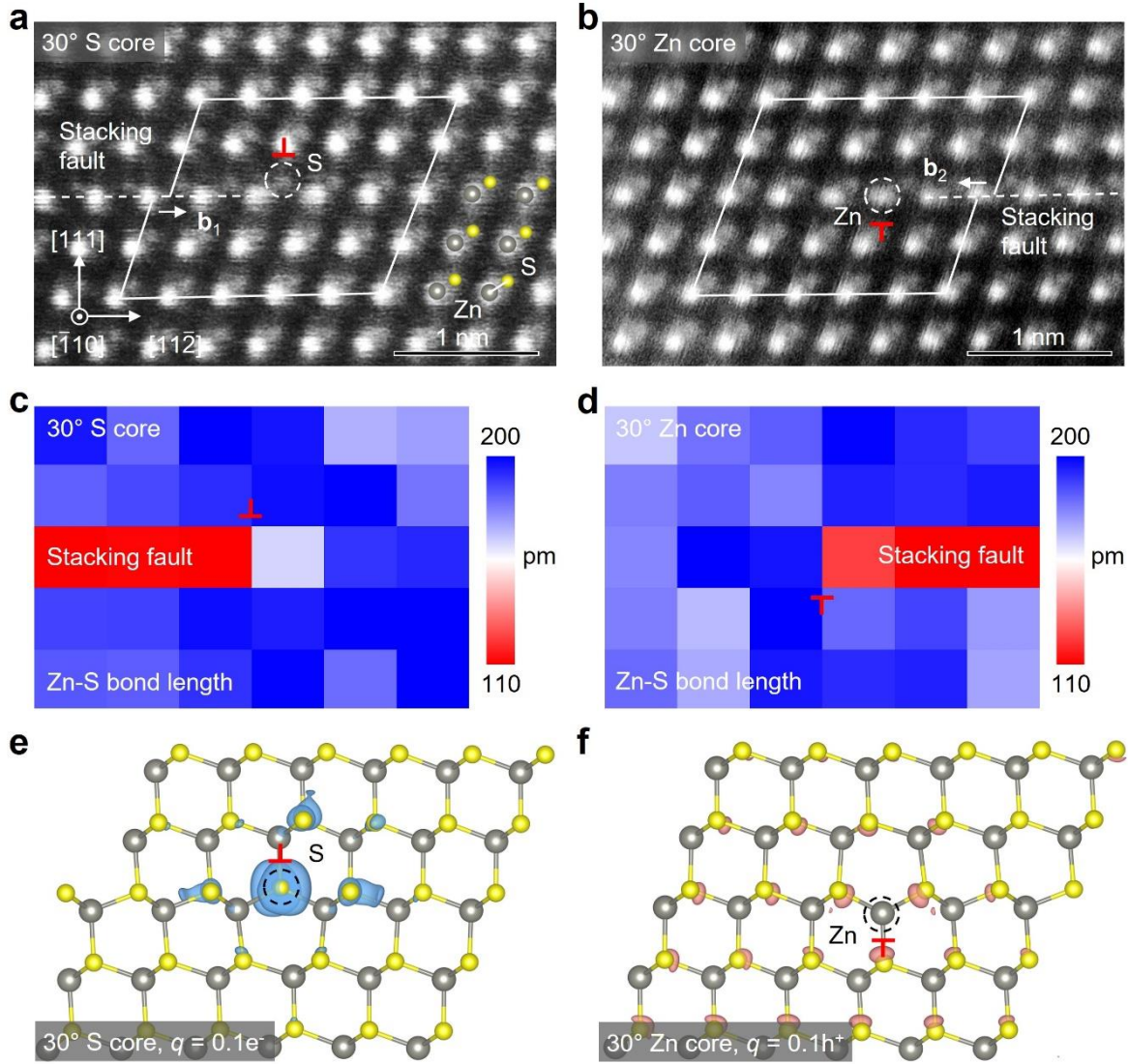

**Fig. 3. Atomic structures and charge distributions of 30° partial dislocations in ZnS.** Atomically resolved HAADF images of 30° S (**a**) and 30° Zn (**b**) dislocation cores, viewed along the direction of the dislocation lines. The Burgers circuits in **a** and **b** show that the projected Burgers vectors $\mathbf{b}_1$ and $\mathbf{b}_2$ are a/12[11$\bar{2}$] and a/12[$\bar{1}\bar{1}$2], respectively (here, a is the lattice parameter of ZnS), indicating that they are 30° partial dislocations (Extended Data Table 1). The white dashed lines indicate the locations of the stacking faults, while the dashed circles indicate terminated elements of the dislocations. **c** and **d**, The Zn-S bond lengths of 30° S (**c**) and 30° Zn (**d**) dislocation cores corresponding to the circled areas in **a** and **b**, respectively. The net charge distributions of the 0.1e⁻ charged 30° S core (**e**) and the 0.1h⁺ charged 30° Zn core (**f**) were obtained from the DFT calculations. The blue and red clouds represent the extra electrons and holes, respectively.

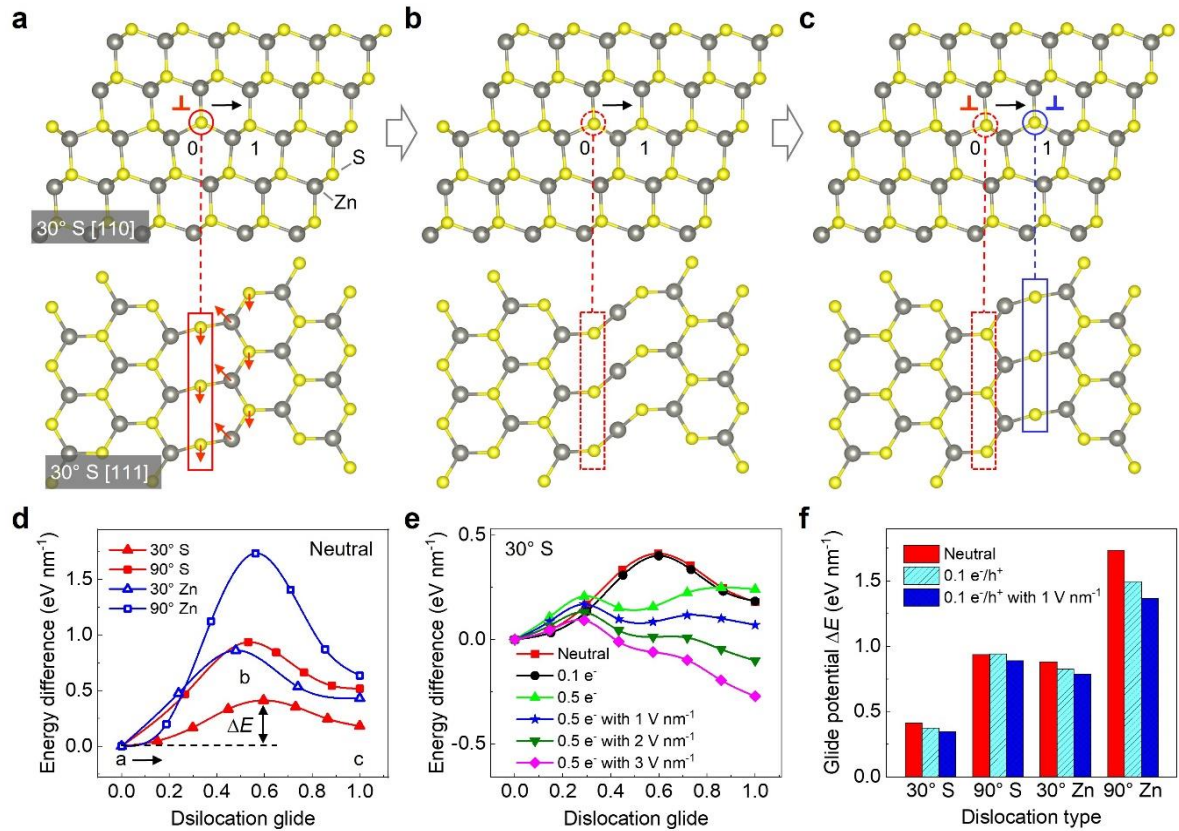

**Fig. 4. Glide barriers of dislocations based on atomic structure evolution. a-c**, The schematics of the initial, transition, and final states of a 30° S core during the glide process, respectively. The circles and rectangles indicate the core positions viewed along the [110] and [111] axes, respectively. The dislocation core is located at Position 0 in the initial state and moves to Position 1 in the final state. The black arrows show the glide direction of the dislocation. The red arrows show the directions of motion of atoms around the dislocation core. **d**, MEPs of the four partial dislocations in their neutral states. The energy of the initial state is set as the reference state, indicated by the dashed line. The glide barrier $\Delta E$ is the energy difference between the initial state and the transition state along the glide path for a unit length of the dislocation line. **e**, MEPs of a 30° S partial dislocation in various charge states and under various electric fields. **f**, The glide barriers of the four partial dislocations in neutral states (red), charged states (cyan), and charged states with an electric field (blue).

## Methods

**Materials.** The single-crystalline ZnS samples are commercial products from Aladdin with a purity of 99.99%. ZnS is a typical inorganic semiconductor with a sphalerite structure ($F\bar{4}3m$) at room temperature. Dislocations in ZnS generally dissociate into partial dislocations due to the low stacking fault energy[52]. A perfect 60° dislocation with **b** = a/2<110> dissociates into a 30° partial dislocation with **b** = a/6<112> and a 90° partial dislocation with **b** = a/6<112>. A screw dislocation dissociates into two 30° partial dislocations with **b** = a/6<112>. The dissociation of 60° dislocations and screw dislocations was observed in this study (Extended Data Fig. 8). The degree of dislocation is defined as the angle between the dislocation line and the Burgers vector. Burger vectors of perfect dislocations and partial dislocations are summarized in Extended Data Table 1. In ZnS, dislocations tend to lie on close-packed {111} planes along <110> directions because the valley of the Peierls potential is along <110> directions [53].

**TEM and STEM characterization.** The TEM samples were prepared using the focused ion beam (FIB) technique (ThermalFisher Helios G4 UX). The ZnS was thinned using a gallium ion beam at 30 kV with a beam current from 49 nA to 7 pA. Lower accelerating voltages 5 kV and 2 kV with a beam current from 20 to 5 pA were used for further thinning to reduce the ion beam damage. Then, a 0.5 - 1 kV ion beam was used for the final cleaning of specimens. Diffraction contrast TEM experiments were carried out using an FEI Tecnai F20 microscope operated at 200 kV. HAADF images were obtained using aberration-corrected FEI Titan Cubed Themis G2 operated at 300 kV. The convergence semi-angle is 30 mrad, and the angular range of the HAADF detector is from 39 mrad to 200 mrad. Positions of Zn and S columns around dislocation cores were determined by simultaneously fitting two-dimensional Gaussian peaks [54].

**In situ TEM characterization.** In situ TEM experiments were carried out using an FEI Tecnai F20 microscope operated at 200 kV with a PicoFemto double-tilt TEM-STM holder from ZEPTools Technology Company. A tungsten tip acts as the top electrode to apply a bias, which was precisely controlled by a piezoelectric system. The voltage range is ±150 V. The copper grid is grounded. The vacuum of the TEM column prevents the breakdown of air under high voltages. During the electrical tests, dislocations are imaged in diffraction contrast by using the reflection **g** = {220}, as labeled in the selected area electron diffraction (SAED) pattern in Supplementary Figs. 2 and Extended Data Fig. 2. Multi beams might be selected for better brightness and contrast of dislocation lines in bright field images. The movement processes were recorded with a OneView camera (Gatan).

**DFT simulations.** We used a prism with a quadrilateral cross-section containing a single dislocation core to perform the first principles simulations. The constructed structures contain one periodic direction of <110>, which is along the dislocation line, and two nonperiodic directions of <111> and <112>. The thickness of the vacuum layer is taken to be about 14 Å to avoid elastic and electrostatic interactions between neighboring prisms in periodic supercells. To passivate dangling bonds at the edge, pseudohydrogen atoms with a fractional charge of 1.5 and 0.5 electrons were terminated to each Zn and S, respectively, to ensure a balanced charge. All the DFT simulations were performed using the VASP package with a plane wave basis set[55, 56]. The generalized gradient approximation (GGA), type Perdew-Burke-Ernzerhof (PBE) functional, was used for exchange and correlation interaction[57, 58]. The projector augmented-wave (PAW) method was used for the core-valence interaction[58], and the $3s^23p^4$ and $3d^{10}4s^2$ electrons were treated as valence states for S and Zn, respectively. The energy cutoff of 350 eV was employed in all simulations. The energy convergence of $10^{-5}$ eV for the electronic self-consistent field (SCF) and the force criterion of $10^{-2}$ eV Å$^{-1}$ were used. A Γ-centered ($1 \times 2 \times 1$) Monkhorst−Pack K-point mesh was used for Brillouin zone integration with a fine resolution less than $2\pi \times 1/40/\text{Å}$[59]. Each dislocation model contains more than 200 atoms. To avoid the influence of the artificial boundaries, we added $0.1e^-$ or $h^+$ in the calculations. We studied unreconstructed dislocation cores and calculated their glide barriers, because the core reconstructions in the zinc-blende structure involve the bonding of second nearest neighbors. However, the ionic bonding in II-VI semiconductors inhibits anion–anion and cation–cation bonding and, thus, makes the core reconstructions in the ionic II-VI semiconductors less favorable compared to elemental IV-IV and III-V covalent semiconductors[60, 61, 62].

**Dislocation structure energy.** The calculated energies of four dislocation structures in different charge states $q$ were calculated using the following equation:

$$E_{cal}(\text{dislocation}, q) = E_{tot}(\text{dislocation}, q) + qE_F$$

where $E_{tot}(\text{dislocation}, q)$ is the total energy of the dislocation structures in different charge states and $E_F$ is the possible Fermi energy for the charged systems, ranging from the valence band maximum (VBM) to conduction band minimum (CBM) in perfect crystal ZnS. Particularly, the zero of the Fermi level corresponds to the valence-band maximum. In the present study, the charge states of $q = 0.1e^-$ or $h^+$ were considered for all four types of dislocations.

**Glide barriers of dislocations.** The climbing image nudged elastic band (NEB) method[43, 44]

was applied to determine energy barriers for the movement of four different types of dislocations in ZnS. The configurations before and after the dislocation movement with a unit vector were used as initial and final configurations in the climbing NEB method. Energy minimization was performed for these initial and final configurations before NEB calculations. During energy minimization, the H atoms terminated at the edge were first fully relaxed while fixing the Zn and S atoms. After that, all the other atoms were relaxed while the H atoms were fixed to mimic the boundary imposed by the rest of the bulk materials. For the work term $\sum_i q_i \mathbf{r}_i \boldsymbol{\epsilon}$, $q_i$ was obtained by the Bader charge analysis[63] and the summation was evaluated and added to the DFT energy for each image along the MEP. Note that only the moving atoms or the redistribution of charges contribute to the changes in the work term between images.

**Simulation of the electric fields.** The distribution of an electric field in the ZnS samples was calculated by using the Ansoft Maxwell software. The tungsten tip and copper grid were assumed as perfect conductors and ZnS as an insulator with a relative dielectric constant of 8.9[64]. The tip diameter and model size were determined by TEM imaging. The tip diameter is about 80 nm in the test of Fig. 1 and 220 nm in the test of Fig. 2. Tip diameters influence the distribution of electric fields. The copper grid is grounded in the experiment. The distributions of electric fields under experimental bias are shown in Supplementary Figs. 1 and 5.

## Data availability

The data that support the findings of this study are available within the article and the Supplementary Information. Any other relevant data are also available upon reasonable request from the corresponding authors.

## Code availability

Additional data including the codes are available from the corresponding authors upon reasonable request.

## Methods-only references

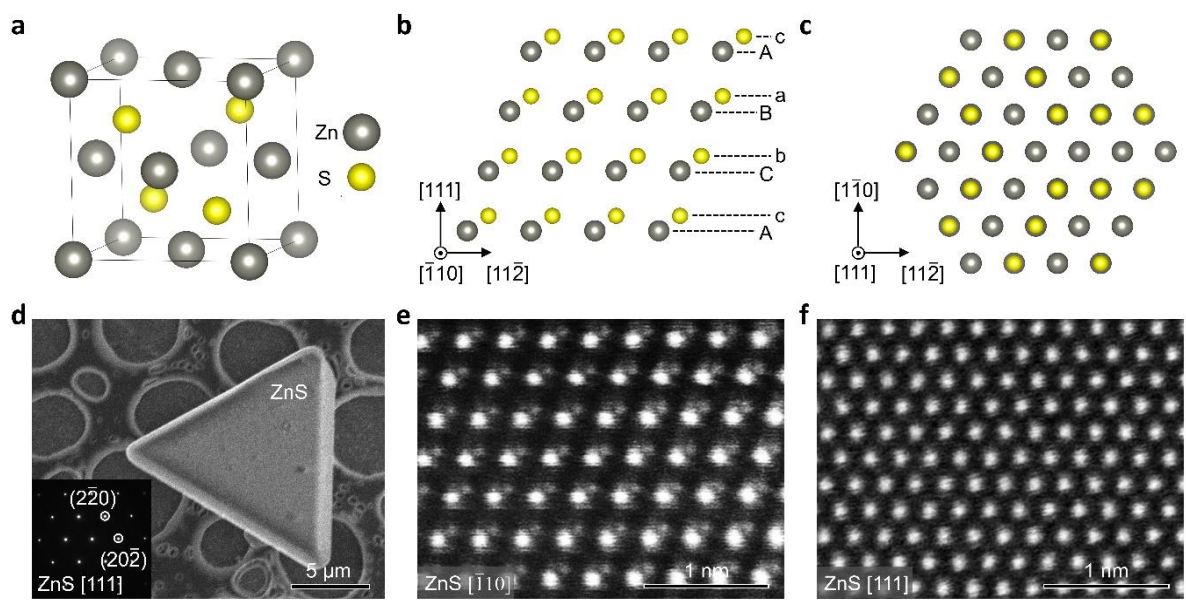

**Extended Data Fig. 1. Microstructure of ZnS samples. a**, Atomic structure of sphalerite ZnS. Projections of sphalerite ZnS along the $[\bar{1}10]$ axis (**b**) and [111] axis (**c**). **d**, Scanning electron microscopy (SEM) image of a ZnS flake with a regular triangular shape. The edge length is about 15 μm. The inset SAED pattern shows the single-crystalline nature of ZnS samples. Atomically resolved HAADF images of the sphalerite ZnS along the $[\bar{1}10]$ axis (**e**) and [111] axis (**f**).

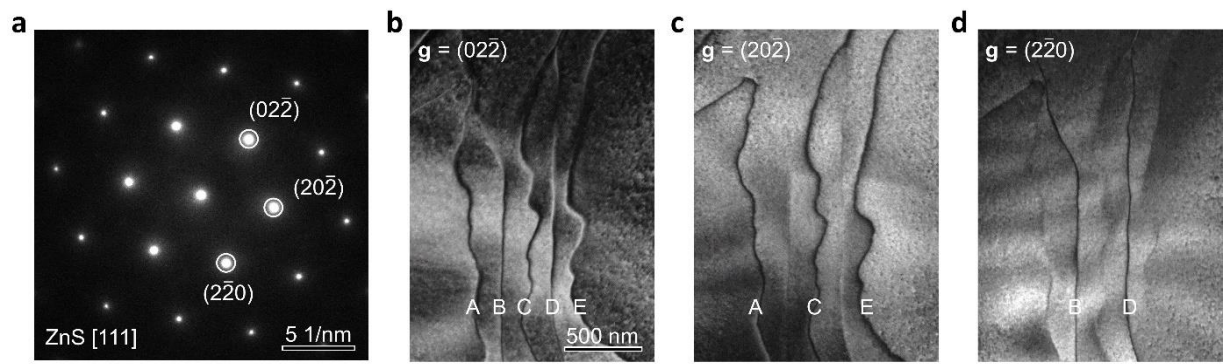

**Extended Data Fig. 2 Characterization of Dislocations A-E in Fig. 2. a**, SAED pattern of the region with five dislocations in Fig. 2. Dark-field TEM images with **g** = ($\bar{2}$02) (**b**), (0$\bar{2}$2) (**c**), (2$\bar{2}$0) (**d**). **g** is the diffraction vector of the operating reflection. Note that Dislocations B and D are invisible in **c** while Dislocations A, C, and E are invisible in **d**. Burgers vectors of dislocations are defined by the invisibility criterion **gb** = 0. Dislocations A, C and E are 90° partial dislocations with the Burgers vector a/6[11$\bar{2}$] while Dislocations B and D are 30° partial dislocations with the Burgers vector a/6[$\bar{2}$11]. Dark points might be amorphous or dust induced by the FIB processing and absorption. The type of dislocation is determined by the angle between the dislocation line and the Burgers vector.

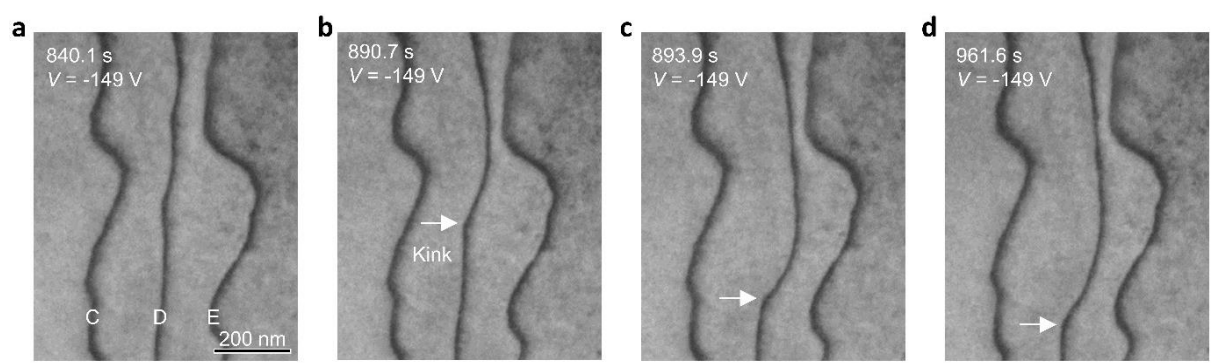

**Extended Data Fig. 3. Pinning (a) and depinning (b) processes during dislocation motion under an electric field.** White circles highlight the positions of two pinning points on Dislocations B and D.

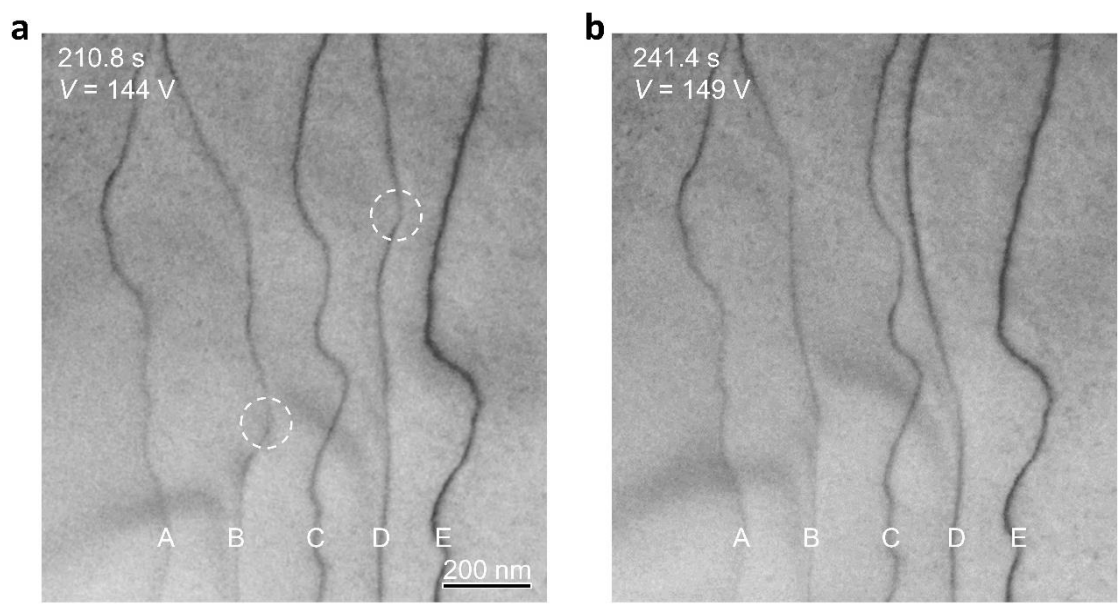

**Extended Data Fig. 4. Kink propagation process. a-d**, Kink propagation along the dislocation line during the in situ electrical testing in Fig. 2. White arrows point to kink positions.

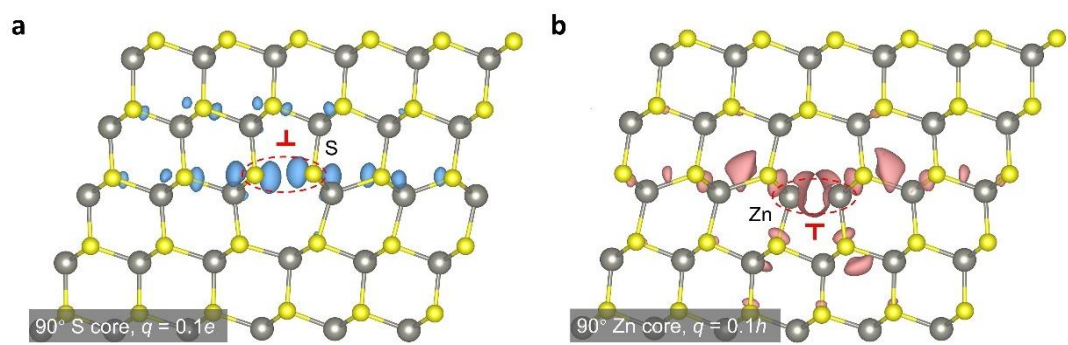

**Extended Data Fig. 5. Net charge distribution of the 0.1e$^-$ charged 90° S core (a) and the 0.1h$^+$ charged 90° Zn core (b) from DFT calculations.** Blue and red clouds represent the extra electrons and holes, respectively.

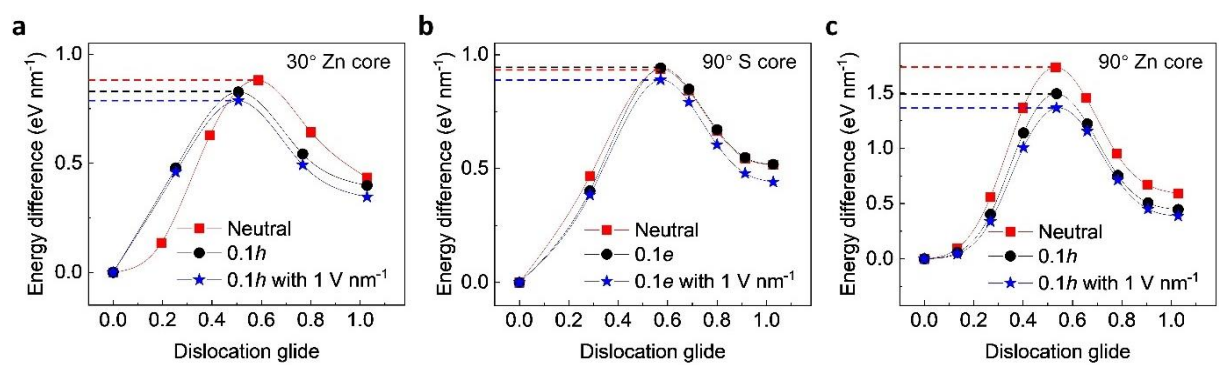

**Extended Data Fig. 6. MEPs of charged dislocations under an applied electric field, including 30° Zn core (a), 90° S core (b), and 90° Zn core (c).**

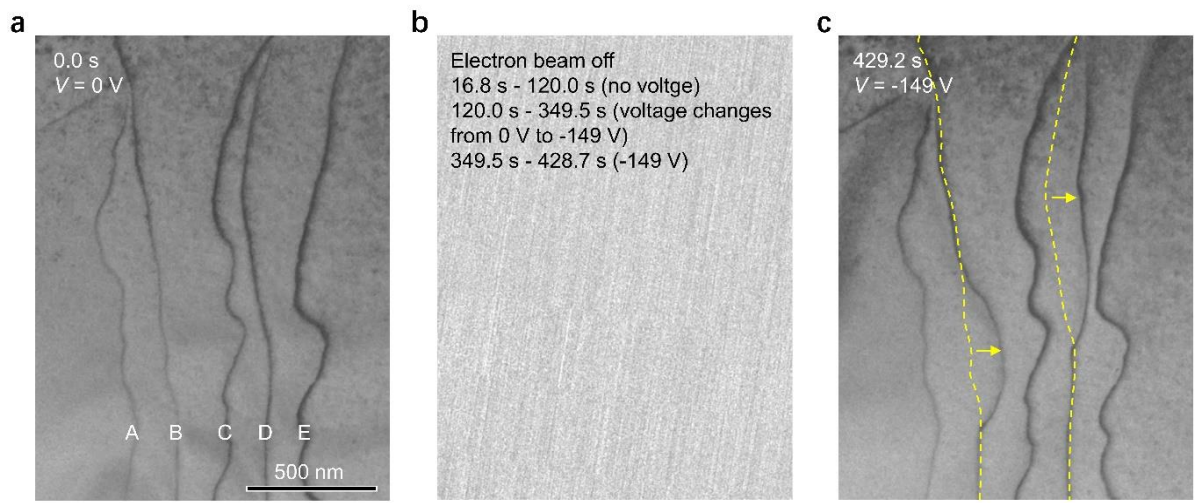

**Extended Data Fig. 7. Dislocation motion driven by an external electric field when the electron beam is off. a,** The TEM image showing the initial positions of five dislocations. **b,** The electron beam is off for about 7 minutes: When the electron beam was off, the applied voltage was changed from 0 V to -149 V and then remained at -149 V for ~1.5 minutes. **c,** The TEM image shows Dislocations B and D moved away from the tip, even with the electron beam off. Dashed lines indicate the initial positions of Dislocation B and D before their motions. Arrows indicate the directions of the dislocation motions.

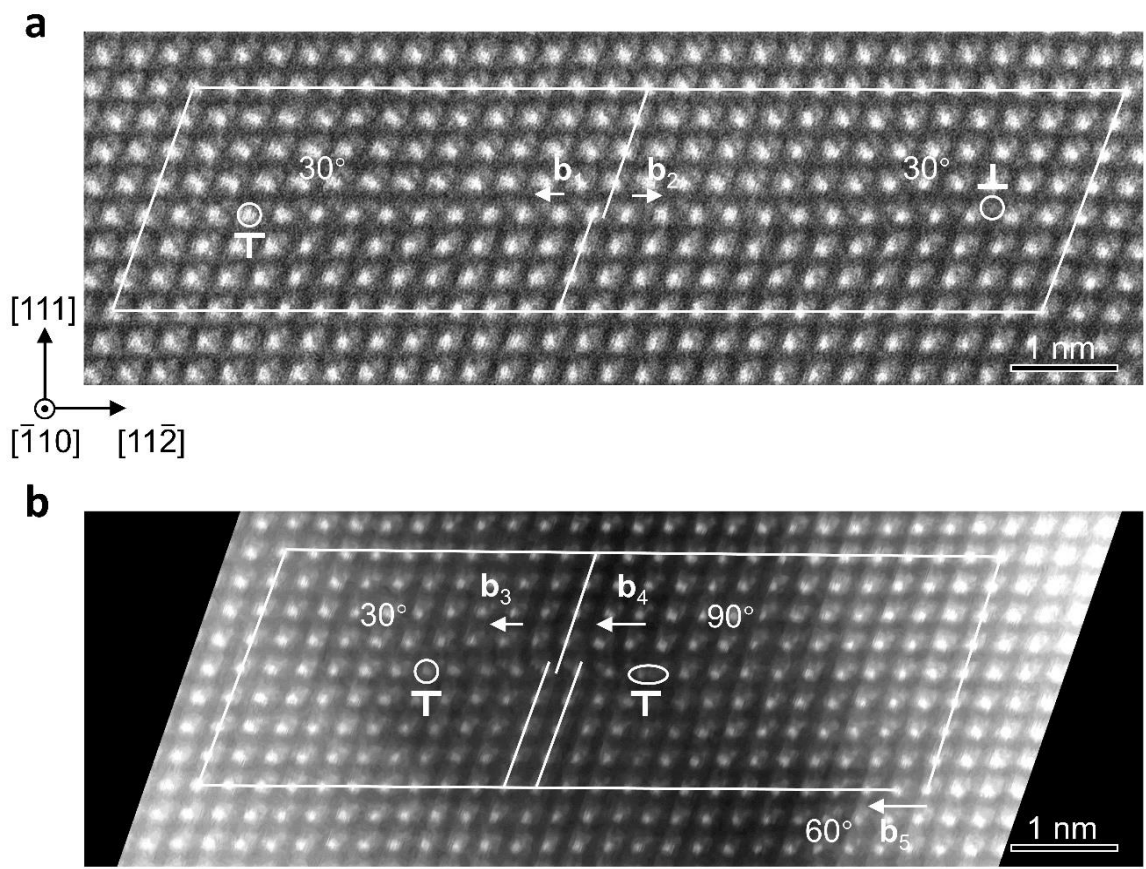

**Extended Data Fig. 8. Dissociation of screw and 60° dislocations in ZnS. a,** HAADF images of a screw dislocation dissociated into two 30° partial dislocations. The closed Burgers circuit shows the overall projected Burgers vectors are zero on the $(\bar{1}10)$ plane, indicating these two partial dislocations are dissociated from a screw dislocation. $b_1$ and $b_2$ indicate the projected Burgers vectors of left and right partial dislocations are $a/12[\bar{1}\bar{1}2]$ and $a/12[11\bar{2}]$, respectively. White circles indicate terminated elements of dislocations cores. This is a stacking fault between two partial dislocations. Scale bar, 1 nm. **b,** HAADF images of a 60° dislocation dissociated into one 30° dislocation and one 90° partial dislocation. $b_3$ and $b_4$ indicate the projected Burgers vectors of left and right partial dislocations are $a/12[\bar{1}\bar{1}2]$ and $a/6[\bar{1}\bar{1}2]$, respectively. The Burgers circuit shows the overall projected Burgers vector $b_5$ is $a/4[\bar{1}\bar{1}2]$ on the $(\bar{1}10)$ plane, indicating these two partial dislocations are dissociated from a 60° dislocation. There is a stacking fault between the two partial dislocations. Scale bar, 1 nm.

|  | Dislocation type | Burgers vector **b** | Projection of **b** on {110} |
|---|---|---|---|
| Perfect dislocations | 60° | a/2<110> | a/4<112> |
|  | Screw | a/2<110> | a/4<112> |
| Partial dislocations | 30° (S or Zn cores) | a/6<112> | a/12<112> |
|  | 90° (S or Zn cores) | a/6<112> | a/6<112> |

**Extended Data Table. 1. Dislocation types and corresponding Burgers vectors in ZnS.**

**This file includes the following:**
Supplementary Figures 1 to 13
Supplementary Tables 1 and 2
Captions for Supplementary Videos 1 to 3

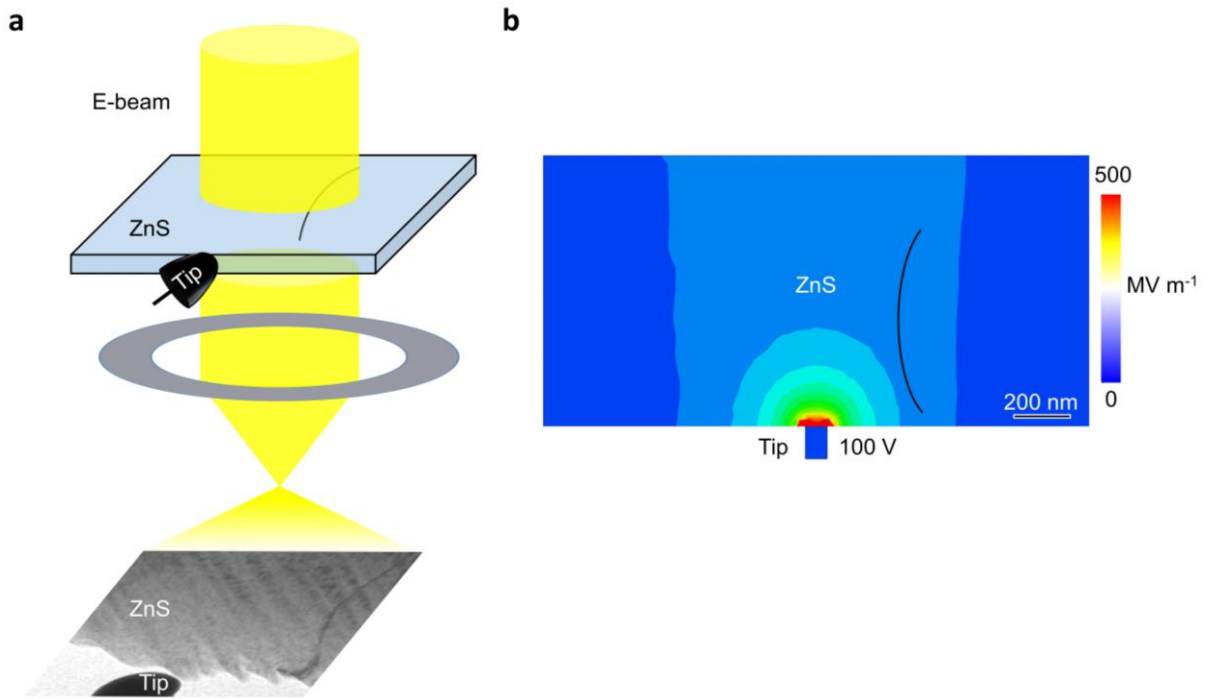

**Supplementary Fig. 1. Electric field distribution of the ZnS sample in Fig.1. a,** Schematic illustration of the experimental set-up in TEM. A voltage (*V*) is applied to ZnS samples through a tungsten tip. **b,** The distribution of the applied electric field in ZnS under 100 V modeled by the Ansoft Maxwell software. The tip contact width is estimated to be 80 nm.

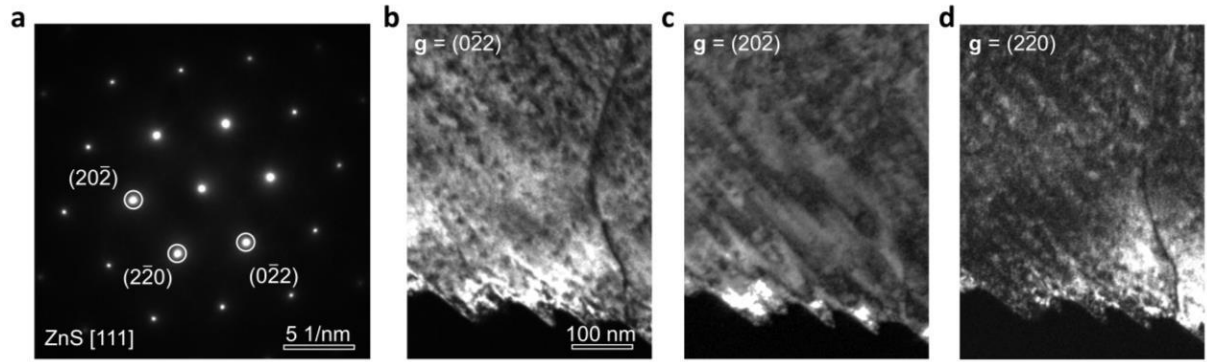

**Supplementary Fig. 2. Characterization of the dislocation in Fig. 1. a**, SAED pattern of the region with the dislocation in Fig. 1. Dark-field TEM images with **g** = $(0\bar{2}2)$ (**b**), $(20\bar{2})$ (**c**), $(2\bar{2}0)$ (**d**). **g** is the diffraction vector of the operating reflection. Note that the dislocation is visible in **b** and **d** while invisible in **c**. The Burgers vectors of the dislocation are defined to be $a/6[1\bar{2}1]$ by the invisibility criterion **gb** = 0. Dark points in TEM images might be amorphous or dust induced by FIB processing and absorption. The dislocation is determined to be a 30° partial dislocation, which is defined as the angle between the dislocation line and the Burgers vector.

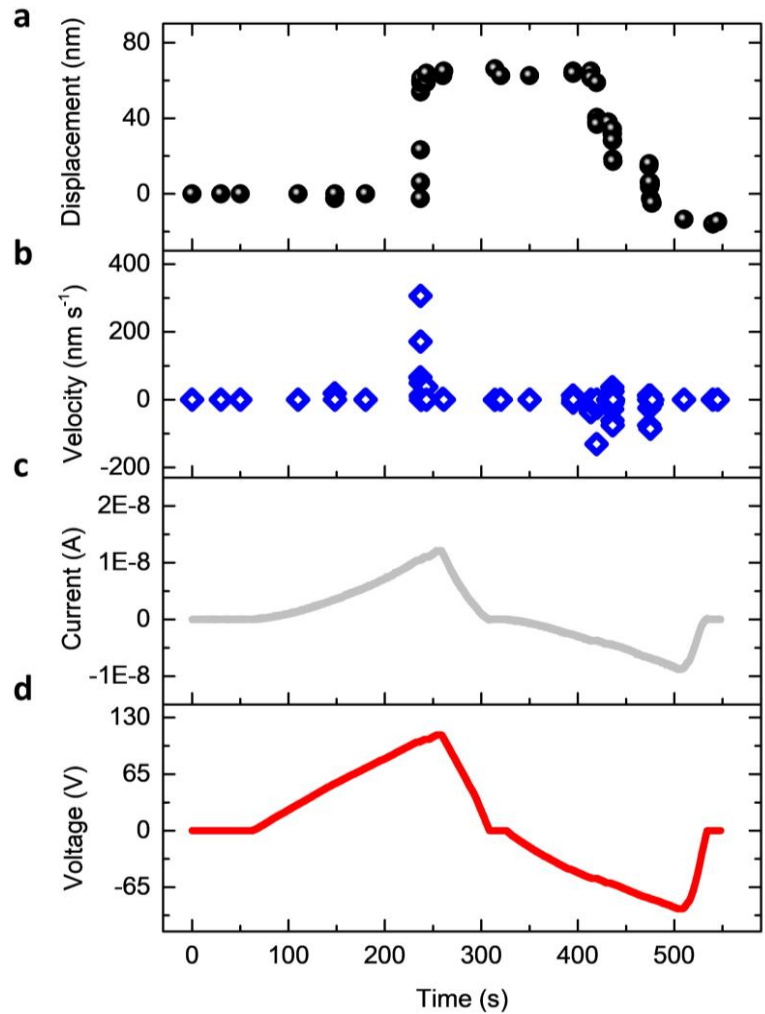

**Supplementary Fig. 3. Measurement of dislocation motion under an electric field in Fig. 1.** Plots of displacement (**a**), velocity (**b**), current (**c**), and applied voltage (**d**) of the marked point in Fig. 1 as functions of time. The displacement range is about 82.1 nm. The maximum velocity is 306.4 nm s$^{-1}$. The point on the dislocation begins to move when voltage is above a threshold value of 51 V, corresponding an electric field of ~$10^7$ V m$^{-1}$. The current is small during tests, ~$10^{-8}$ A. The voltage increases slowly, about 1 V s$^{-1}$, to avoid thermal strain induced by Joule heating.

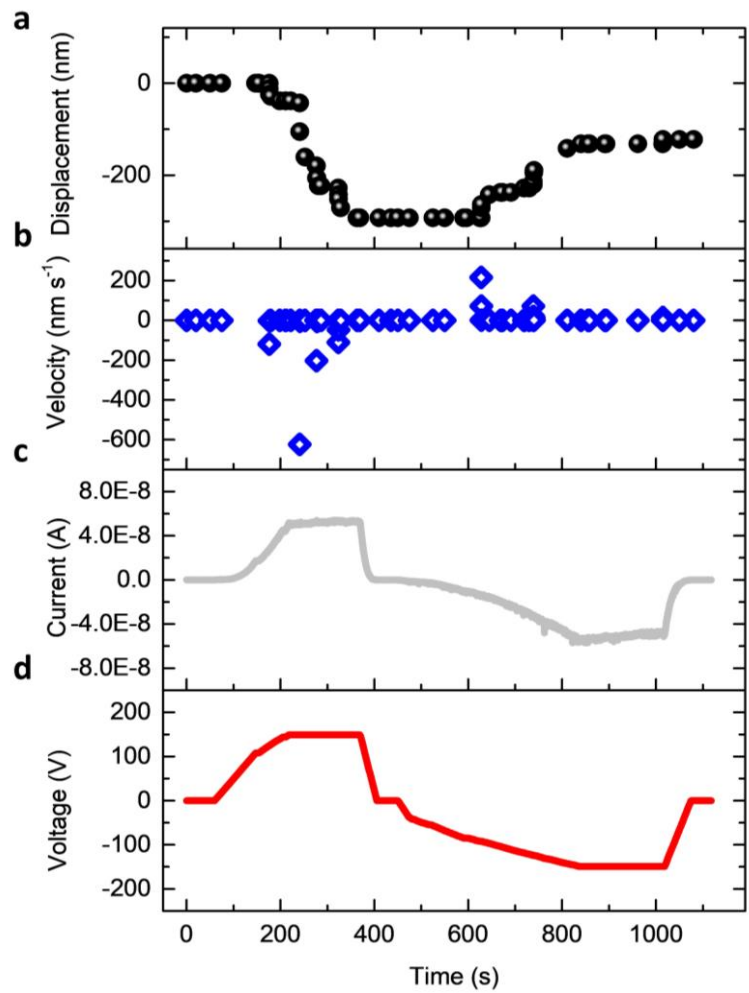

**Supplementary Fig. 4. Measurement of dislocation motion under electric fields in Fig. 2.** Plots of applied voltage (**a**), current (**b**), displacement (**c**), and velocity (**d**) of marked points on Dislocation B as functions of time. The displacement range is about 292.6 nm. The maximum velocity appears at high voltages in **a**, up to 623.5 nm s⁻¹. The point on the dislocation begins to move when the voltage is above a threshold value of 104 V. The current is very small during tests, below $8×10^{-8}$ A. The voltage increases slowly, about 1 V s⁻¹, to avoid thermal strain induced by Joule heating.

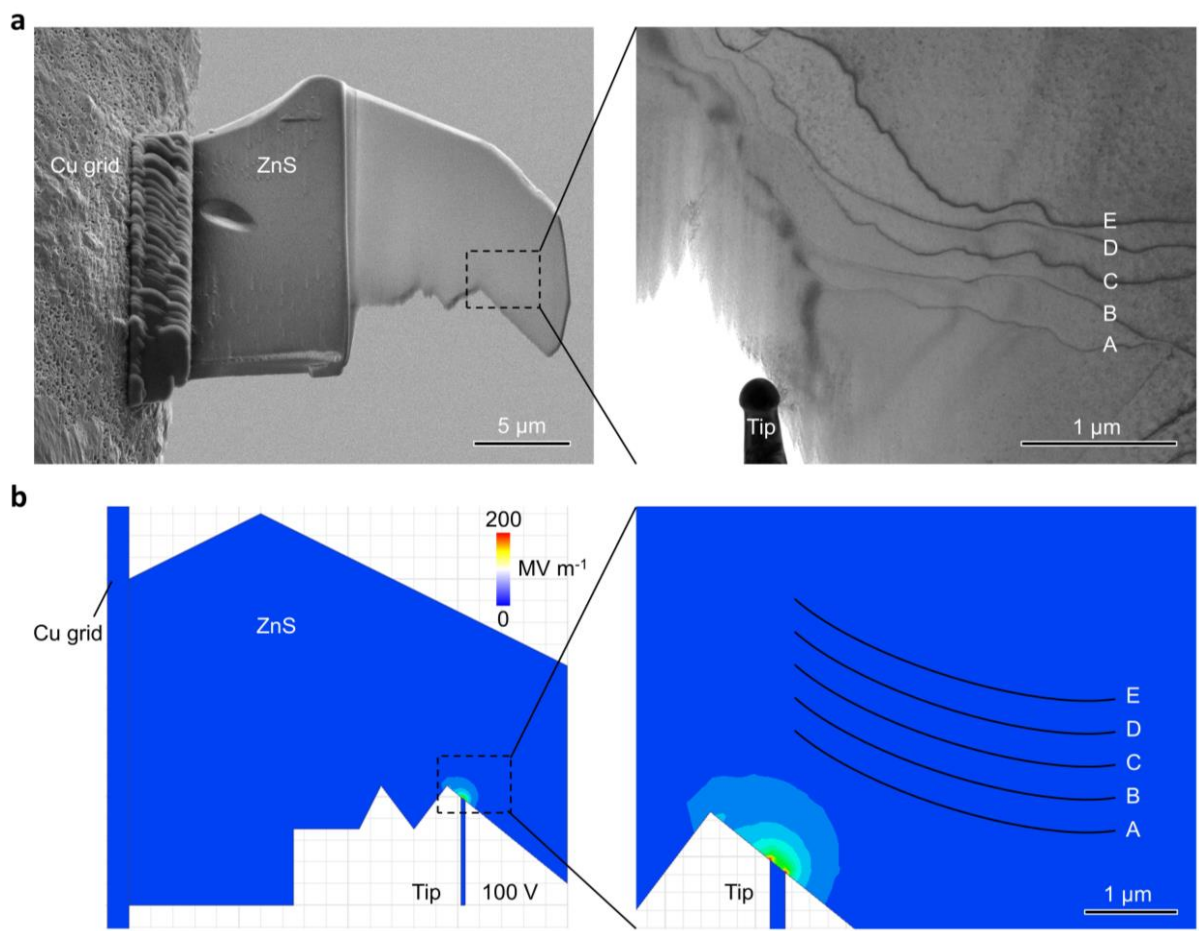

**Supplementary Fig. 5. The distribution of the electric field of the ZnS sample in Fig. 2. a**, SEM image of the ZnS sample in Fig. 2. The enlarged TEM image shows positions of dislocations and the tip with a scale bar 1 μm. The Cu grid is grounded. **b**, The distribution of the applied electric field in ZnS under 100 V modeled by the Ansoft Maxwell software. The tip contact width is estimated to be 220 nm. The enlarged image shows the electric field around dislocations is about $10^6 \sim 10^7$ V m$^{-1}$.

**Supplementary Table 1. Starting electric field of marked points in Fig. 2.**

|  | Towards tip (V m$^{-1}$) | Away from tip (V m$^{-1}$) |
|---|---|---|
| Marked point on Dislocation B | $6.9\times10^{6}$ | $-5.6\times10^{6}$ |
| Marked point on Dislocation D | $5.4\times10^{6}$ | $-5.1\times10^{6}$ |

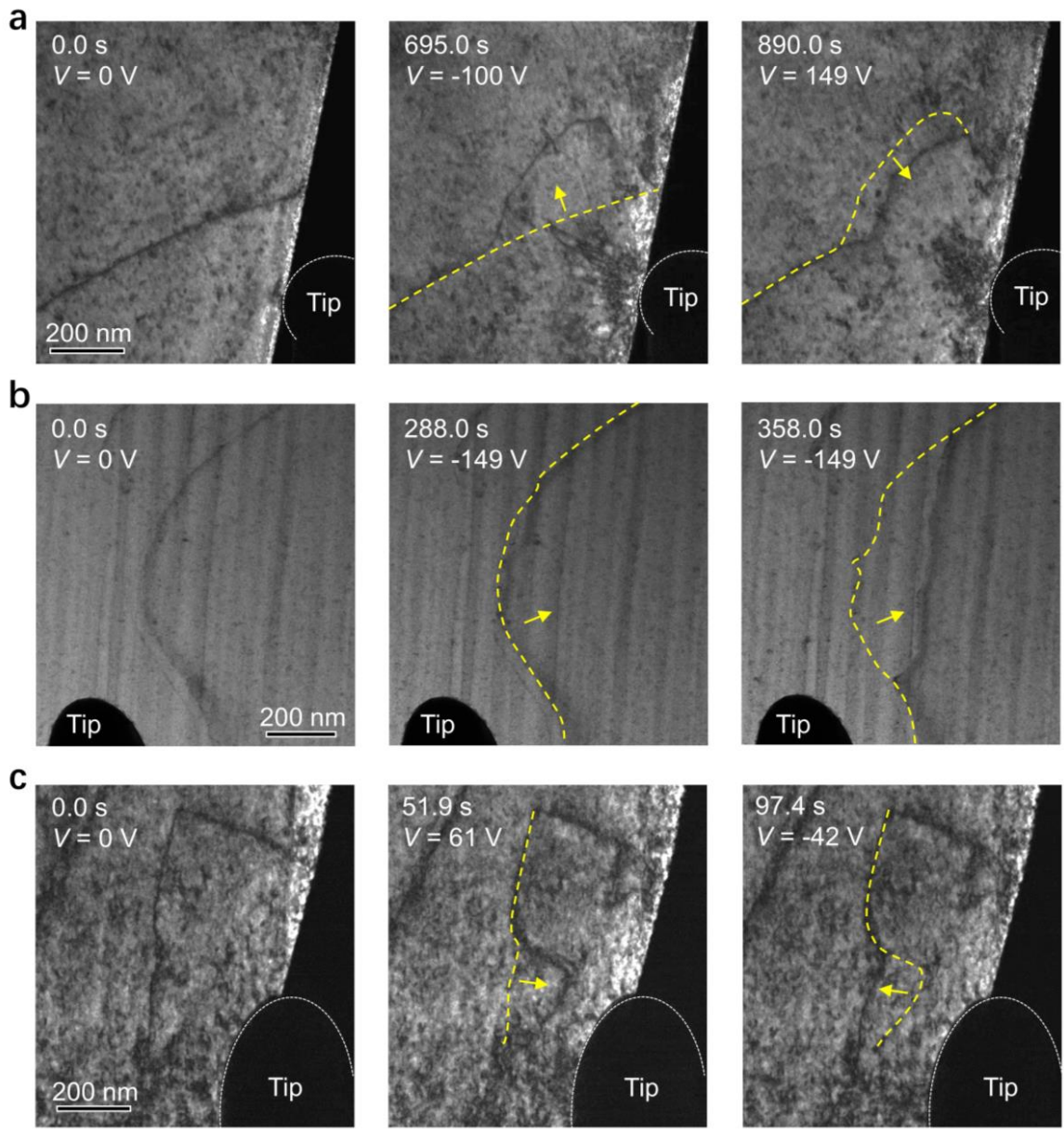

**Supplementary Fig. 6. a-c, Three additional representative experimental results demonstrating the phenomenon of electrical-driven dislocation motion.** Dashed lines indicate the initial positions of the dislocations. Arrows show the directions of the dislocation motion.

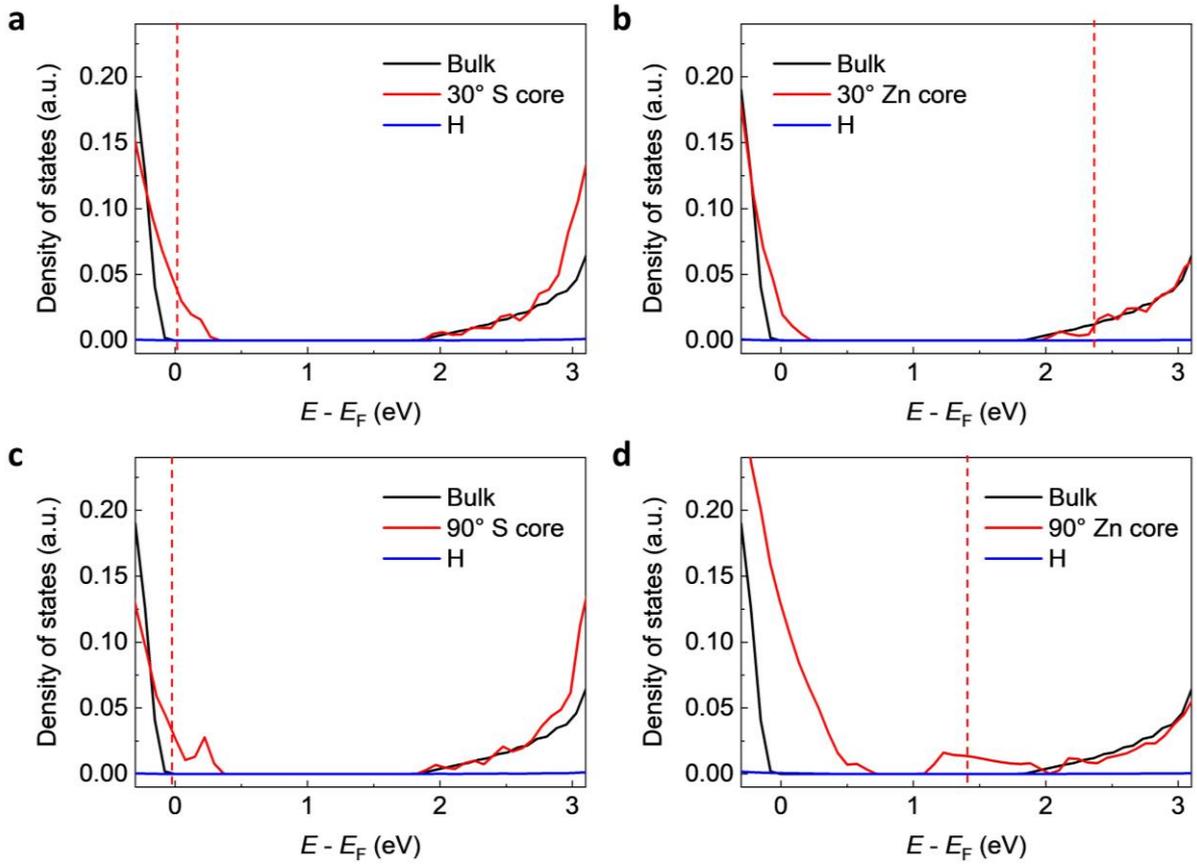

**Supplementary Fig. 7**. **Density of states of partial dislocations in ZnS. a**, 30° S core. **b**, 30° Zn core. **c**, 90° S core. **d**, 90° Zn core. The bulk Fermi level (valance band maximum) is set to zero. The red dashed lines indicate the Fermi level of the dislocation cores. The Fermi level of dislocations is 0.075 eV for 30° S core, 2.371 eV for 30° Zn core, -0.006 eV for 90° S core, and 1.428 eV for 90° Zn core. Blue lines indicate the DOS of terminating H atoms, which is negligible. The DOS of the bulk is included as a reference.

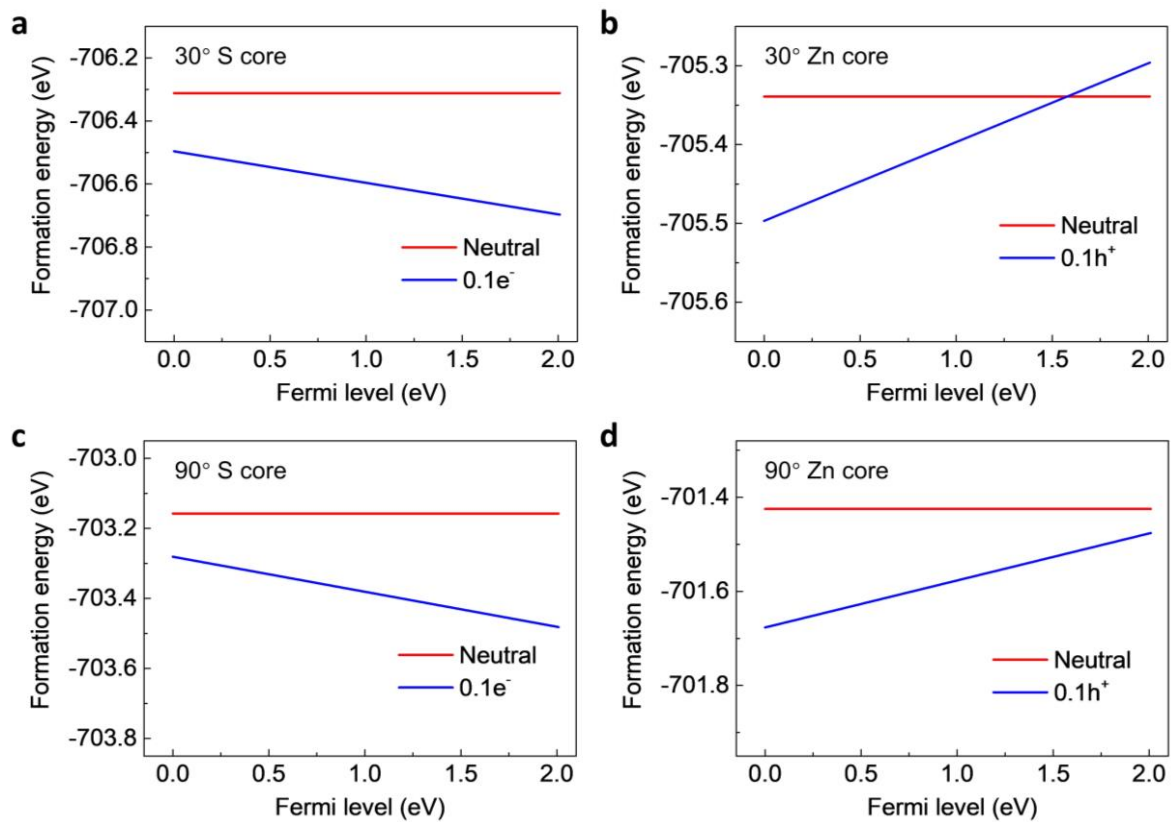

**Supplementary Fig. 8. Dislocation structure energy as a function of the Fermi level of the system in different charge states in ZnS. a**, 30° S core. **b**, 30° Zn core **c**, 90° S core **d**, 90° Zn core. The zero of the Fermi level corresponds to the bulk valence band maximum.

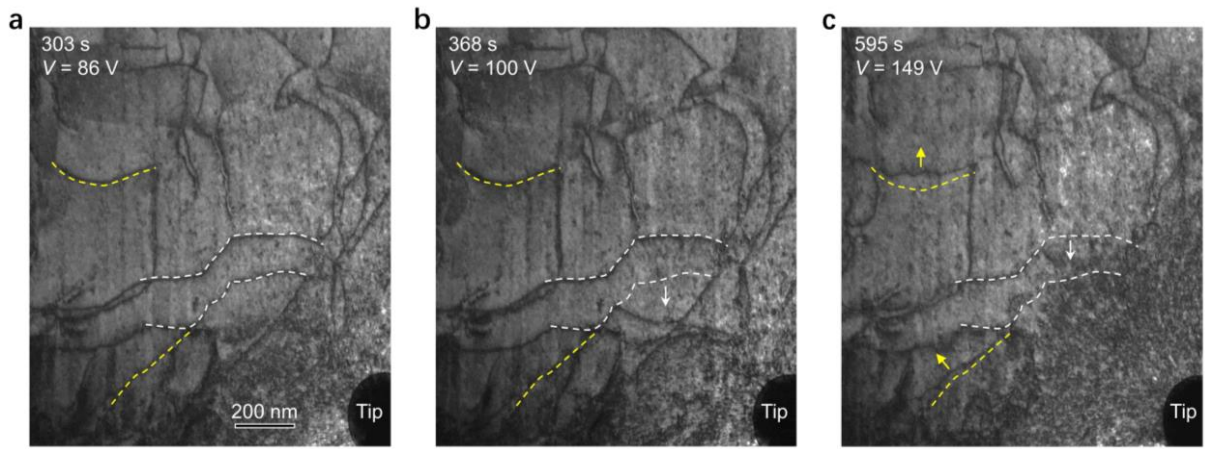

**Supplementary Fig. 9. The opposite motion directions of dislocations observed in the same specimen under an electric field. a-c,** The dislocations marked by yellow arrows move away from the tip, while the dislocations marked by white arrows move towards the tip. Dashed lines indicate the initial positions of the dislocations.

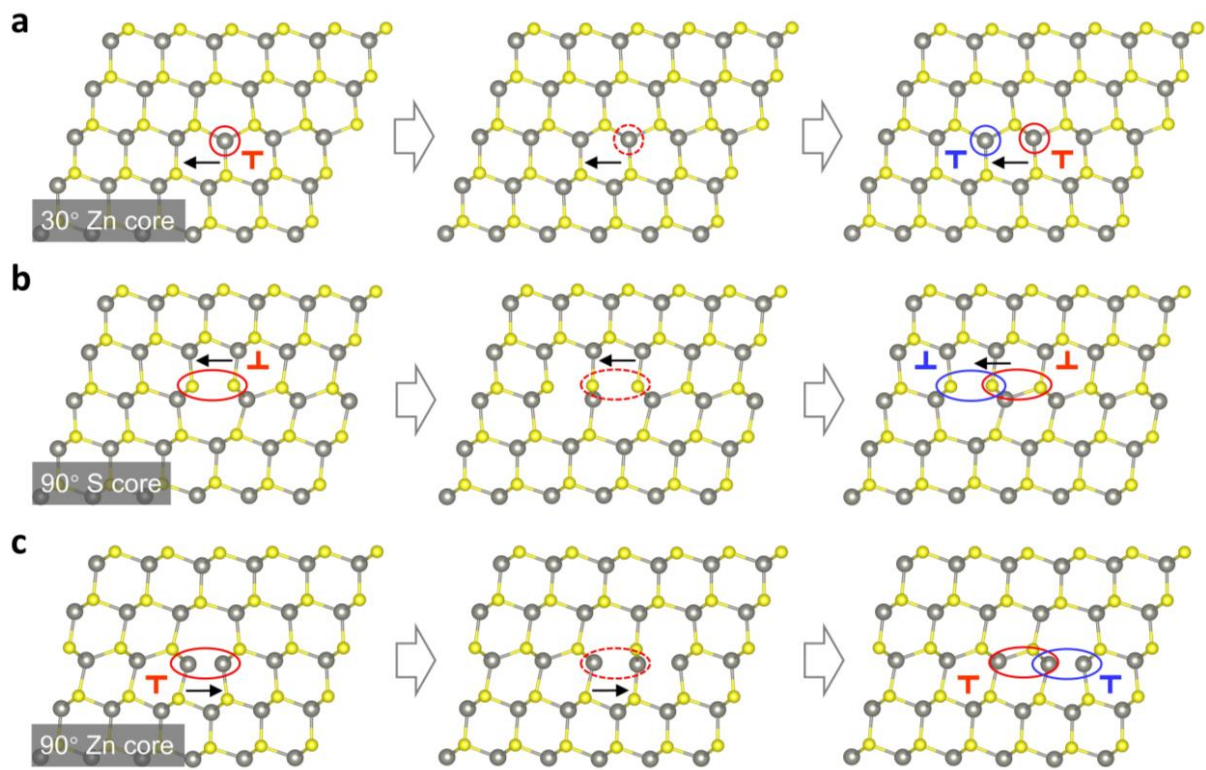

**Supplementary Fig. 10. Glide of dislocations based on the atomic structure evolution of dislocation cores. a**, Typical states of a 30° Zn dislocation core during glide. The circles mark core atoms. **b**, Typical states of a 90° S dislocation core during glide. **c**, Typical states of a 90° Zn dislocation core during glide.

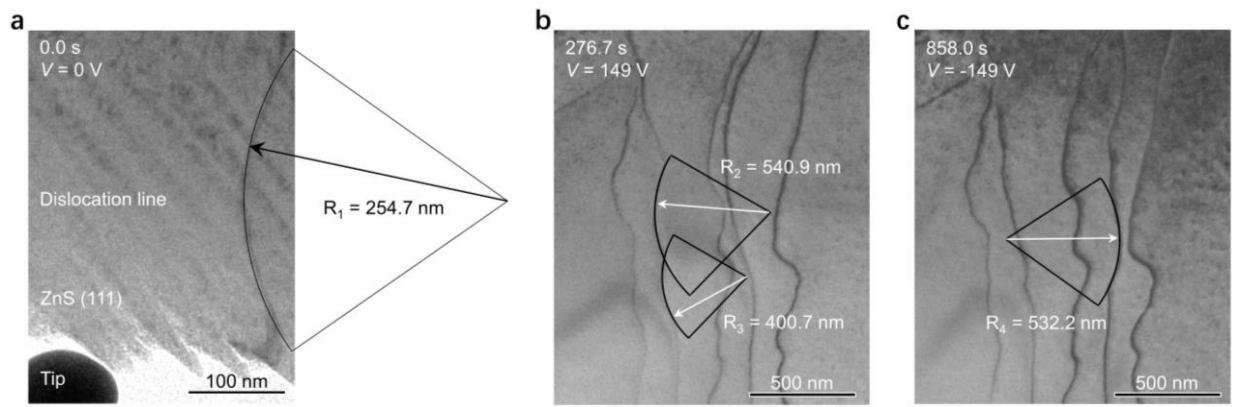

**Supplementary Fig. 11. The curvatures of the dislocation lines in Figs. 1 and 2.** (a) The measured radius of the curvature of the dislocation segment in Fig. 1 is 254.7 nm. (b) and (c) The measured radii of the curvatures are 540.9 nm, 400.7 nm and 532.2 nm for three dislocation segments in Fig. 2, respectively.

**Supplementary Table 2. The estimated critical resolved shear stress ($\tau$) and electrostatic stress on the dislocation** segments **in Figs. 1 and 2.**

| Dislocation segments | Radius of curvatures ($R$) | $\tau_1 = \frac{Gb}{2R}$ | $\tau_2 = \frac{Gb}{4\pi R} \times \frac{2-\nu}{2(1-\nu)} \ln \frac{4R}{eb}$ [2] | $\tau_\epsilon = \frac{\epsilon\rho}{b}$ [3] |
|---|---|---|---|---|
| In Fig.1 | $R_1 = 254.7$ nm | 13.22 MPa | 19.33 MPa | 22.63 MPa ($\epsilon = 2.4 \times 10^7$ V m$^{-1}$) |
| In Fig.2 | $R_2 = 540.9$ nm | 6.23 MPa | 10.03 MPa | 5.47 MPa ($\epsilon = 5.8 \times 10^6$ V m$^{-1}$) |
| | $R_3 = 400.7$ nm | 8.41 MPa | 13.04 MPa | |
| | $R_4 = 532.2$ nm | 6.33 MPa | 10.17 MPa | |

where $G$ is the average shear modulus of ZnS ~30.5 GPa[4], **b** the Burgers vector of the 30° dislocation **b** = a/6<112>, $R$ the radius of the curved dislocation segments (Supplementary Fig. 17), $\nu$ the Poisson ration, $\nu = 0.32$[2], $\epsilon$ the critical electric field to move the dislocations in our experiments. In Fig. 1, $\epsilon = 2.4 \times 10^7$ V m$^{-1}$ in Fig. 2, $\epsilon = 5.8 \times 10^6$ V m$^{-1}$. $\rho$ is the charge density of the dislocation, which is assumed to be 1.3 e$^-$ nm$^{-1}$ (0.5 e$^-$ on the 0.385 nm dislocation line used in our atomic model, according to Fig. 4e).

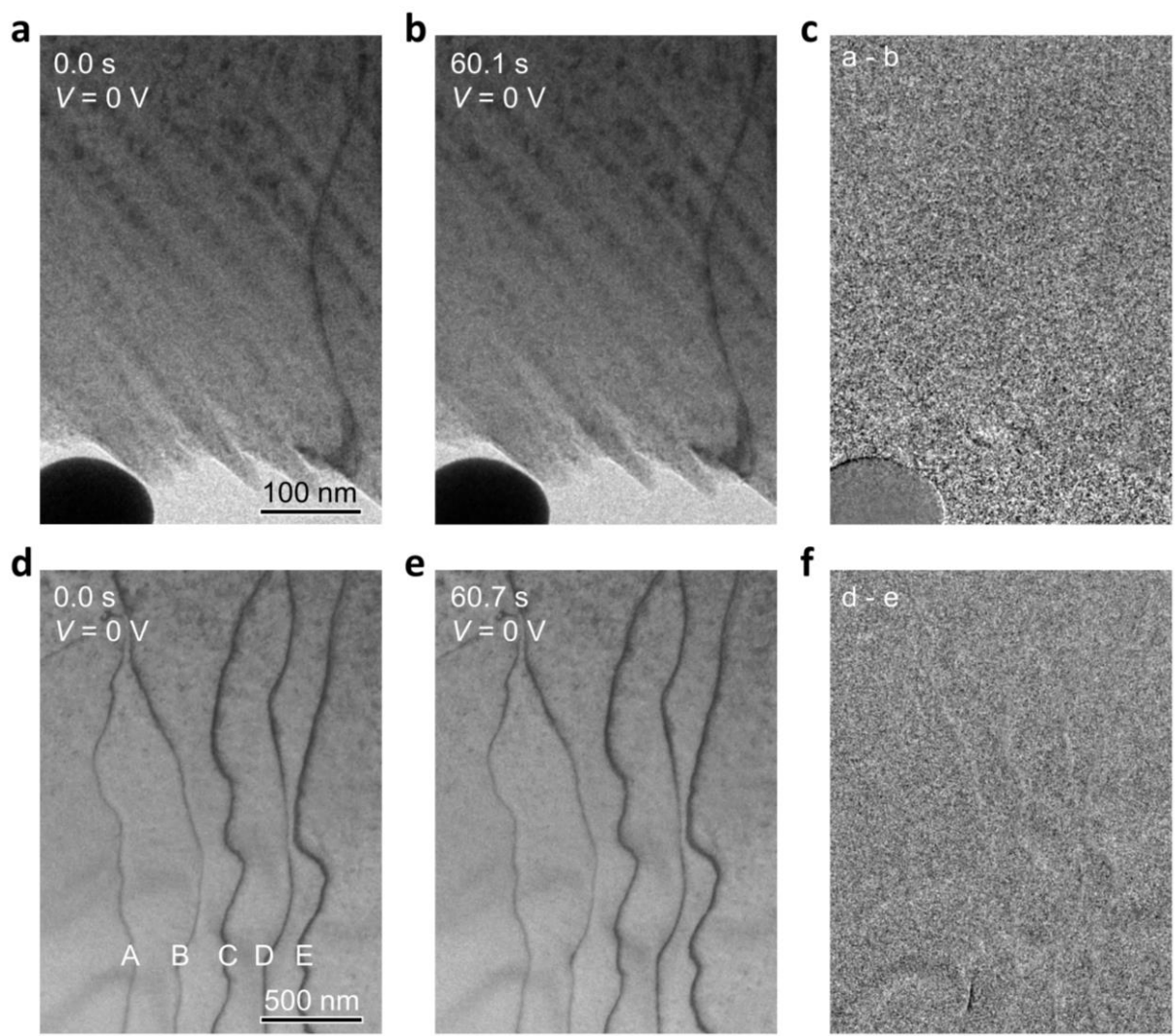

**Supplementary Fig. 12. The effect of electron beam irradiation on dislocations. a**, Initial state of the dislocation in Fig. 1. **b**, The same region after 60.1 s under the electron beam irradiation. **c**, The contrast from **a** minus **b**. The movement of the dislocation will induce a contrast change in **c**. **d**, the Initial state of the dislocations in Fig. 2. **e**, The same region after 60.7 s under the electron beam irradiation. **f**, The contrast from **d** minus **e**. The movement of dislocations will induce a contrast change in **f**. No reversible motion was observed under electron beam irradiation.

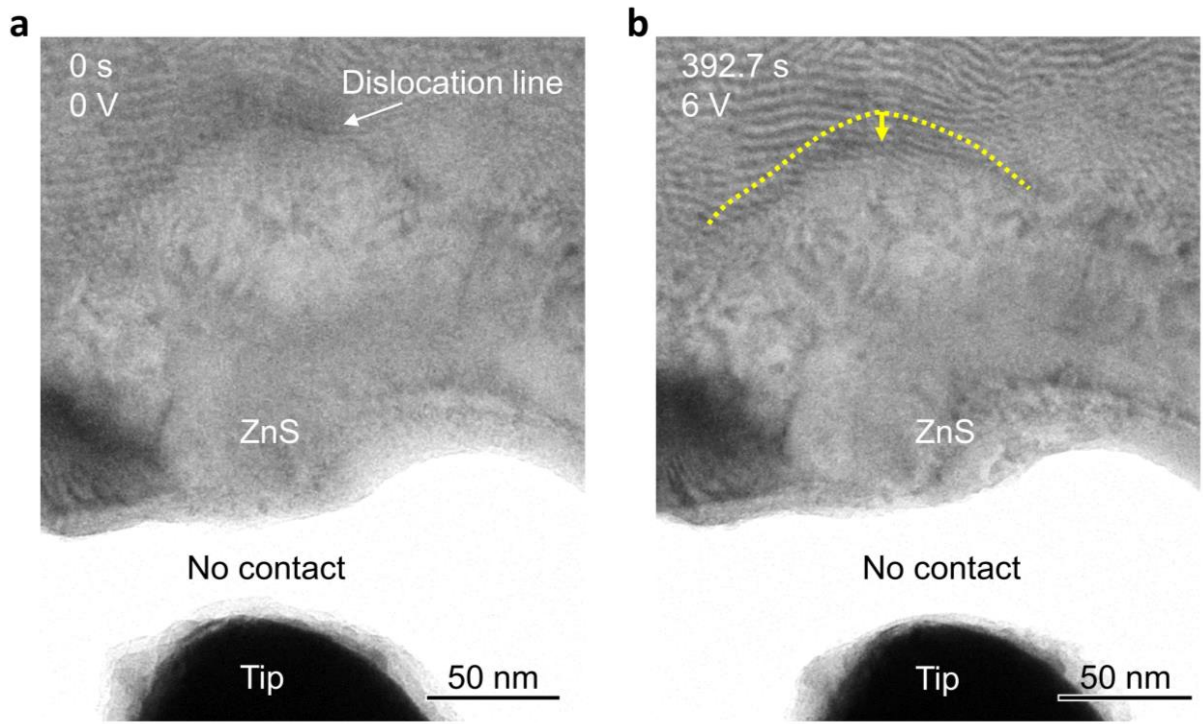

**Supplementary Fig. 13. Driving dislocation motion by an electric field with no contact with a specimen. a,** The initial state of the dislocation. **b,** The dislocation moved to a new position under an applied electric field.

**Captions for Videos**

**Supplementary Video 1**. In situ TEM observation of an individual dislocation moved in an external electric field. The dislocation moves back and forth depending on the magnitude and direction of the electric field. From 0 s to 260 s, the applied voltage increases from 0 V to +110 V. The dislocation moves away from the tip. From 260 s to 320 s, the applied voltage decreases from +110 V to 0 V and the dislocation does not move. From 320 s to 510 s, the applied voltage changes from 0 V to -90 V, and the dislocation moves in an opposite direction (towards the tip). From 510 s to the end of the video, the applied voltage returns to 0 V and the dislocation remains at its position.

**Supplementary Video 2**. The different mobility of 30° and 90° partial dislocations in an electric field. The 30° partial dislocations (Dislocations B and D in Fig. 2) move back and forth, driven by the applied electric field while the 90° partial dislocations (Dislocations A, C, and E in Fig. 2) are motionless during the entire experiment. From 0 s to 370 s, the applied voltage increases from 0 V to +149 V. Two 30° partial dislocations move towards the tip. From 370 s to 450 s, the applied voltage decreases from +149 V to 0 V and the dislocations do not move. From 450 s to 1010 s, the applied voltage changes from 0 V to -149 V and the two 30° partial dislocations move in an opposite direction (away from the tip). From 1010 s to the end of the video, the applied voltage returns to 0 V and the dislocations remain at their positions. The three 90° partial dislocations are motionless during the entire test.

**Supplementary Video 3**. Dislocation motion driven by an external electric field in the condition the electron beam is off. To eliminate the dominate effect of electron beam irradiation on the dislocation motion, an additional test was carried out in the same region of Fig. 2 with the electron beam off. After recording the initial state of dislocations, the electron beam was turned off, after ~2 minutes we applied a voltage up to -149 V to the sample. The voltage was remained at the peak value of -149 V or ~1.5 minutes. And then, we turned on the electron beam and recorded the positions of dislocations again. By comparing the TEM images, we found that Dislocations B and D moved to the right even when the electron beam was off.